# Detection of the Minimal Supersymmetric Model Higgs Boson H⁰ in its h⁰h⁰ → 4b and A⁰A⁰ → 4b Decay Channels

J. Dai[a], J.F. Gunion[a], and R. Vega[c]

[a] *Physics Department, University of California, La Jolla, CA 92093, USA*

[b] *Davis Institute for High Energy Physics, University of California, Davis, CA 95616, USA*

[c] *Physics Department, Southern Methodist University, Dallas, TX 75275*

## Abstract

We demonstrate that detection of the heavier minimal supersymmetric model CP-even Higgs boson $H^0$ will be possible at the LHC via its $H^0 \to h^0 h^0 \to 4b$ and/or $H^0 \to A^0 A^0 \to 4b$ decay channels for significant portions of the $(m_{A^0}, \tan\beta)$ model parameter space. At low $m_{A^0}$ ($\lesssim 60$ GeV), *both* the $H^0 \to A^0 A^0 \to 4b$ and $H^0 \to h^0 h^0 \to 4b$ modes yield a viable signal for most $\tan\beta$ values; viability for the $h^0 h^0$ channel extends up to $m_{H^0} \sim 2 m_t$ when the model parameter $\tan\beta$ is not large. At the Tevatron, the $h^0 h^0$ and $A^0 A^0$ channels are both potentially viable at low $m_{A^0}$ for sufficiently good $b$-tagging efficiency and purity.

## I. INTRODUCTION

The minimal supersymmetric standard model (MSSM) is currently the most attractive theory for physics below the TeV energy scale. The two Higgs doublets of the model ($H_1$ and $H_2$) are precisely the number required for: i) anomaly cancellations; ii) giving masses to both up quarks ($H_2$) and down quarks and leptons ($H_1$); and iii) nearly precise gauge coupling constant unification. While extra singlet Higgs fields can be added without affecting these crucial features, extra doublets or any number of Higgs triplet fields would destroy unification. Thus, it is vital to assess our ability to fully explore the Higgs sector of the MSSM at future colliders.

The physical Higgs bosons of the MSSM are: two CP-even Higgs boson, the $h^0$ and $H^0$ (with $m_{h^0} < m_{H^0}$ by definition); one CP-odd Higgs boson, the $A^0$; and a pair of charged Higgs bosons, $H^\pm$. The $h^0$ will very probably be detected at either LEP II or the LHC. However, the $h^0$ might well have properties very like those of the single Higgs boson $h_{SM}$ of the minimal standard model (SM). The easiest, and perhaps only, way in which to verify that the Higgs sector is non-minimal would then be detection of a second Higgs boson. Unfortunately, our ability to do so at either the LHC or a future NLC is far from guaranteed. (The extensive literature on this subject is reviewed in Ref. [1].) Further, it would be highly desirable to be able to test the very explicit predictions for the self-couplings among the Higgs bosons of the MSSM. In this article, we demonstrate that, for the expected levels of $b$ tagging efficiency and purity, the production/decay mode $gg \to H^0 \to h^0 h^0 \to b\bar{b}b\bar{b}$ will be observable at the LHC in important regions of the parameter space of the model, thereby allowing both detection of the $H^0$ and sensitivity to the $H^0 h^0 h^0$ coupling. We also show that when kinematically allowed the $H^0 \to A^0 A^0 \to 4b$ channel will be observable for most $\tan\beta$ values. In addition, we explore the observability of these modes at an upgraded Tevatron with integrated luminosity in the vicinity of $L = 30$ fb$^{-1}$, and find statistically significant signals at small $m_{A^0}$, provided the efficiency and purity of $b$ tagging are both excellent.

To begin, we note some key features of the MSSM Higgs sector. At tree-level, the Higgs sector is completely determined by the two parameters $m_{A^0}$ and $\tan\beta = v_2/v_1$ (where $v_1$ and $v_2$ are the vacuum expectation values of the neutral components of the $H_1$ and $H_2$ doublets). After including radiative corrections at one-loop [2] and two-loops [3,4], other parameters of the model, especially the top-quark (pole) mass $m_t$ and the squark masses and mixings, have a strong influence on the Higgs sector. At large $m_{A^0}$ ($\gtrsim$ 150 GeV), $m_{h^0}$ approaches an upper bound, while at small $m_{A^0}$, $m_{H^0}$ approaches a lower bound, and both these bounds increase with increasing $m_t$ and $m_{\tilde{t}}$. For $m_{A^0}$ between 20 and 70-80 GeV, $m_{h^0} \sim m_{A^0}$ when $\tan\beta \gtrsim 4 - 6$.

For all the calculations of this paper we shall take $m_t = 175$ GeV, $m_{\tilde{t}} = 1$ TeV and neglect squark mixing. For these choices, one finds that $m_{h^0}$ is sufficiently small, even after radiative corrections, that the $H^0 \to h^0 h^0$ decay channel is essentially always open for small to moderate $\tan\beta$, although strongly kinematically suppressed for $m_{A^0}$ in the vicinity of $\sim 90$ GeV. For larger $\tan\beta$ values, the $H^0 \to h^0 h^0$ decay is allowed only for $m_{A^0} \lesssim 60$ GeV and $m_{A^0} \gtrsim 200$ GeV. For $m_{A^0} \lesssim 60$ GeV, the $H^0 \to A^0 A^0$ channel is also kinematically allowed. These kinematics, in combination with the complicated coupling patterns of the Higgs bosons, yield the $H^0 \to h^0 h^0$ branching ratio contours plotted in Fig. 1 in $(m_{A^0}, \tan\beta)$ parameter space. Full two-loop corrections to the $H^0$ and $h^0$ masses, mixing angles, and (very importantly) Higgs self-couplings have been included. We see that even though $H^0 \to h^0 h^0$ is generally kinematically allowed for $m_{A^0} \gtrsim 200$ GeV, $BR(H^0 \to h^0 h^0)$ is not guaranteed to be large; it is suppressed at large $\tan\beta$ values by the dominance of the enhanced $H^0 \to b\overline{b}, \tau^+\tau^-$ modes and for $m_{H^0} \sim m_{A^0} \gtrsim 2m_t$ by the dominance of $t\overline{t}$ decays. In addition, we note that the $H^0 \to A^0 A^0$ mode generally has a sizeable branching ratio whenever it is kinematically allowed, *i.e.* for $m_{A^0} \lesssim 60$ GeV. Finally, the decays of the $h^0$ are essentially always dominated by $b\overline{b}$, as are those of the $A^0$ when $m_{A^0}$ is small.

These results assume that SUSY decay channels are absent or negligible. When kinematically allowed, SUSY decays will be important for small to moderate $\tan\beta$ values and would suppress $BR(H^0 \to h^0 h^0)$.

## II. LHC RESULTS

At the LHC, the channels that have been discussed for detection of the $H^0$ at moderate $\tan\beta$ and $m_{A^0} \lesssim 2m_t$ are $gg \to H^0$ production followed by (a) $H^0 \to ZZ^{(*)} \to 4\ell$, (b) $H^0 \to h^0 h^0 \to b\overline{b}b\overline{b}$ and (c) $H^0 \to h^0 h^0 \to b\bar{b}\gamma\gamma$, the inclusive $H^0 \to b\bar{b}$ channel having far too large a QCD background. The $4\ell$ channel (a) has been the object of numerous theoretical and ATLAS+CMS experimental studies [5,1]. It has been shown to be viable for (roughly) 120 GeV $\lesssim m_{A^0} \lesssim 2m_t, \tan\beta \lesssim 1.5$ at low luminosity ($L = 30$ fb$^{-1}$) and 70 GeV $\lesssim m_{A^0} \lesssim 2m_t, \tan\beta \lesssim 3-5$ at high luminosity ($L = 300$ fb$^{-1}$). ATLAS and CMS have recently explored the $b\bar{b}\gamma\gamma$ mode (c). At $L = 300$ fb$^{-1}$ it is viable when $BR(H^0 \to h^0 h^0) \gtrsim 0.05$ (see Fig. 1) and $m_{A^0} \gtrsim 120$ GeV [5,1]. The $b\bar{b}\gamma\gamma$ channel has small backgrounds, but the expected number of events is quite limited since the $h^0 \to \gamma\gamma$ branching ratio is of order $10^{-3}$ for much of parameter space. Here, we study whether the much higher rates in the $H^0 \to h^0 h^0 \to 4b$ channel can be utilized to detect the $H^0 \to h^0 h^0$ decays. In conjunction with the $2b2\gamma$ channel, this would also allow a model independent determination of the very important ratio $BR(h^0 \to \gamma\gamma)/BR(h^0 \to b\bar{b})$. In what follows, we demonstrate that the $4b$ channel will, indeed, allow detection of the $H^0$ in its $h^0 h^0$ decay mode in several important regions of parameter space, including a large region which overlaps that where the $2b2\gamma$ mode is viable. In addition, we find that, when kinematically allowed, the $H^0 \to A^0 A^0 \to 4b$ channel will be detectable.

We will consider three $b$-tagging scenarios. The first is what now seems to be a relatively conservative scenario taken from the SDC detector Technical Design Report [7] and employed in our earlier work [6]. As a function of $p_T$, $e_{b-tag}$ takes the values 0.16 at $p_T = 20$ GeV, 0.22 at $p_T = 30$ GeV, and is $\geq 0.3$ for large $p_T$ (yielding a rough average value of $e_{b-tag} \sim 0.25$); the probability of mis-tagging is taken to be $e_{mis-tag} = 0.01$ for all $p_T \geq 20$ GeV. For this $b$-tagging scenario we assume a maximal yearly integrated luminosity of $L = 100$ fb$^{-1}$. Current expectations further improve the situation. Canonical values now employed by ATLAS and CMS [5] are: $e_{b-tag} \sim 0.6$ and $e_{mis-tag} \sim 0.01$ for $|\eta| \leq 2.5$ and $p_T \geq 15$ GeV at low luminosity (applicable for accumulated luminosity of $L = 30$ fb$^{-1}$); and $e_{b-tag} \sim 0.5$ and $e_{mis-tag} \sim 0.02$ for $|\eta| \leq 2.5$ and $p_T \geq 30$ GeV at high luminosity ($L = 300$ fb$^{-1}$ accumulated). In the present work, we shall examine all three of these $b$-tagging scenarios; we shall label them (I), (II) and (III), respectively, in the order of the above discussion.

In this paper, we focus on $gg \to H^0$ production followed by $H^0 \to h^0 h^0 \to 4b$ or $H^0 \to A^0 A^0 \to$

$4b$ decay. The $gg \to b\bar{b}H^0$ cross section is not significantly enhanced in the regions of parameter space where $BR(H^0 \to h^0 h^0)$ is substantial (see Fig. 1), and, in fact, is small by comparison to the inclusive $gg \to H^0$ cross section in these regions. Although there is no special enhancement for the $H^0$ production rate, the $gg \to H^0 \to h^0 h^0 \to 4b$ ($gg \to H^0 \to A^0 A^0 \to 4b$) channels have the advantage over the previously-studied $gg \to b\bar{b}H^0 \to 4b$ channel that we can require two $2b$ combinations of mass $\sim m_{h^0}$ ($\sim m_{A^0}$) as well as a peak in the $4b$ mass. This considerably reduces both combinatoric and QCD backgrounds. Assuming 3 or 4 $b$-tagging, we find that the only possibly important backgrounds to the $H^0 \to h^0 h^0 \to 4b$ process are the irreducible $gg \to b\bar{b}b\bar{b}$ background and the mis-tag $gg \to b\bar{b}gg$ background (the intrinsic rate for this latter being very large). Because of the large size of $m_t$, the rate for $t\bar{t}b\bar{b}$ production is sufficiently smaller than that for $b\bar{b}b\bar{b}$ as to be neglectable, and the rate for $gg \to b\bar{b}q\bar{q}$ (summed over all light $q \neq b$) is much smaller than that for $gg \to b\bar{b}gg$. (We compute signal and background processes using exact parton level matrix elements.)

For the original conservative $b$-tagging scenario (I) it is difficult to maintain adequate event rates if we demand that 4 jets be tagged as $b$'s. Thus, in this case, we shall only require that 3 *or* 4 jets be tagged. The largest background is then from $b\bar{b}gg$ where at least one of the $g$'s is mis-tagged. For $b$-tagging scenarios (II) and (III) $e_{b-tag}$ is sufficiently large that requiring 4 jets to be tagged as $b$'s leaves an adequate signal rate while suppressing the potentially large $gg \to b\bar{b}gg$ background to a level below the irreducible $gg \to b\bar{b}b\bar{b}$ background.

Semi-leptonic decays of the $b$-jets are included in our analysis. To incorporate detector resolution effects, we smear the lepton momenta using $\delta E/E = 0.2/\sqrt{E(GeV)} + 0.01$ and the quark momenta using $\delta E/E = 0.5/\sqrt{E(GeV)} + 0.03$. If a $b$-jet decays semi-leptonically, we replace the jet momentum by the sum of the (smeared) lepton and light quark momenta; otherwise, we employ the full (smeared) $b$-jet momentum. Approximate techniques for compensating for the momentum lost to the neutrino in the case of a semi-leptonic decay have not been employed here; implementation of such a technique might improve our results.

The precise cuts imposed are adjusted according to the (presumed) masses of the $H^0$ and $h^0$ (or $A^0$). In the following we use the generic notation $h$ for the $h^0$ or light $A^0$. First, we required that there be two jet pairs each having pair mass in the interval $[0.88 m_h, 1.08 m_h]$. Second, we required that the $4b$ mass be in an appropriate interval: $[115, 135]$, $[140, 160]$, $[160, 185]$, $[180, 210]$, $[235, 260]$, $[285, 310]$, $[330, 360]$, $[380, 415]$, $[430, 465]$, and $[475, 515]$ for $m_{H^0}$ = 125, 150, 175, 200, 250, 300, 350, 400, 450, and 500, respectively (all masses in GeV). These mass intervals in $m_h$ and $m_{H^0}$ are roughly optimal for the smearing outlined above. They extend further below the nominal mass than above so as to capture more of the low mass tail resulting from semi-leptonic $b$ decays.

We will also impose a cut requiring that the transverse missing energy be less than 35 GeV. This cut does not significantly reduce the signal rate since signal events with substantial missing energy will not pass the $m_h$ cuts anyway. This is because there are typically several semi-leptonic $b$ decays in events with large $\not{E}_T$, making it quite likely that one or more of the pairs of $b$ quarks that come from the decay of the $h$'s will not reconstruct to a pair mass close to $m_h$. In contrast, the $\not{E}_T \leq 35$ GeV cut removes the substantial percentage of background events where the semi-leptonic $b$ decays lead to large missing energy. We note that in computing the missing energy and determining the optimal cut, we have included an estimate of the missing energy associated with the underlying minimum-bias event structure.

For each choice of $m_{H^0}$ and $m_h$ we considered two possible $p_T^{\min}$ cut options and two possible cuts on the minimum separation required between any two of the $b$ jets, $\Delta R^{\min}$. These were $p_T^{\min}$ = 15 or 30 GeV and $\Delta R^{\min}$ = 0.7 or 1.2. (Note that the $p_T^{\min}$ = 15 GeV cut cannot be employed in the high luminosity scenario III, and that for scenario I the effective minimum is 20 GeV.) For any given combination of $m_{H^0}$, $m_h$ masses the choice among these four ($\Delta R^{\min}, p_T^{\min}$) options which gives the largest significance for the signal is chosen. Generally speaking, it is advantageous to impose the strongest cuts that one can without harming the efficiency of acceptance too severely. Roughly, for lower values of $m_{H^0} \lesssim 200$ GeV, the lower $p_T^{\min}$ must be employed, whereas for $m_{H^0} \gtrsim 200$ GeV the higher $p_T^{\min}$ was generally more optimal. When $m_{H^0}$ is large and $m_h$ is small, the $h$ boost in the $H^0$ center of mass is often such that the smaller $\Delta R^{\min}$ must be employed.

Our procedure was to tabulate the signal and background rates for a series of $m_{H^0}$ masses and $m_h$ masses that scanned all $m_{H^0}$ values of possible interest and all $m_h$ values for which $H^0 \to hh$ is allowed by kinematics, for each of the four $(\Delta R^{\min}, p_T^{\min})$ options described above. (Of course, only the two $\Delta R^{\min}$ options were examined in the high-luminosity $b$-tagging case (III) for which $p_T^{\min} = 30$ GeV is required.) Rates corresponding to the best of these options for each $m_{H^0}, m_h$ (in terms of statistical significance) were then entered into a final summary table. SUSY relations between $m_{H^0}$ and $m_h$ were not implemented in constructing the tables. Signal event rates for $gg \to H^0 \to hh \to 4b$ were obtained assuming 100% branching ratio for the decays $H^0 \to hh$ and $h \to b\overline{b}$ and assuming a width $\Gamma(H^0 \to gg) = 5 \times 10^{-4}$ GeV, that is typical of the values obtained in the MSSM. In order to determine the actual statistical significance of the $gg \to H^0 \to h^0 h^0 \to 4b$ and $gg \to H^0 \to A^0 A^0 \to 4b$ signals as a function of location in the MSSM parameter space, rates were obtained for the appropriate $m_{H^0}$ and $m_h = m_{h^0}$ (or $m_h = m_{A^0}$) values by interpolation between entries in the summary table, and the branching ratios and $gg$ width were adjusted according to the predicted MSSM values. In our considerations, $m_h$ masses below 20 GeV are not considered nor are $m_{H^0}$ masses below 110 GeV. The contours presented below correspond to setting statistical significances to zero for such cases.

It is not appropriate here to detail the signal and background rates tabulated in the summary table. We confine ourselves to a few comments.

1. $b$-tagging option (I): 3 or 4 $b$-tags, $L = 100$ fb$^{-1}$ — In this case, after cuts and tagging the $gg \to H^0 \to 4b$ signal and $gg \to 4b$ background rates are typically comparable, whereas the $gg \to 2b2g$ mis-tag background rate is a factor of 3-9 higher.

2. $b$-tagging option (II): low luminosity ($L = 30$ fb$^{-1}$), 4 $b$-tags — Here, the $gg \to H^0 \to 4b$ rate is somewhat larger than the $gg \to 4b$ irreducible background rate, which, in turn, is a factor of at least 10 larger than the $bb \to 2b2g$ rate — i.e. the latter background becomes unimportant when 4$b$'s are tagged with excellent efficiency and purity.

3. $b$-tagging option (III): high luminosity ($L = 300$ fb$^{-1}$), 4 $b$-tags — The $p_T^{\min} = 30$ GeV rates are obtained from those in option (II) by multiplying the rates for 4$b$ final state channels by the factor $(0.5/0.6)^4 \times 10 \simeq 4.8$ and the rates for the $2b2g$ background by $(0.5/0.6)^2 \times (0.02/0.01)^2 \times 10 \simeq 27.8$. In this case, the $2b2g$ background rates are typically comparable to the $4b$ background rates.

The relation between the option (I) and option (II) rates is easily understood. Very roughly, the signal and $gg \to 4b$ rates for option (I) are

$$\propto \left\{ 4 \times (e_{b-tag} \sim 0.25)^3 \times [(1 - e_{b-tag}) \sim 0.75] + (e_{b-tag} \sim 0.25)^4 \right\} \times 100 \text{ fb}^{-1} = 5.1 \text{ fb}^{-1}$$

(the factor of 4 corresponding to any one of the 4 $b$'s remaining untagged), while those for option (II) are $\propto (e_{b-tag} = 0.6)^4 \times 30 \text{ fb}^{-1} = 3.9 \text{ fb}^{-1}$. In contrast, in going from option (I) to option (II) the $gg \to 2b2g$ rates are suppressed by a factor of roughly

$$\frac{(0.6)^2 \times (0.01)^2 \times 30 \text{ fb}^{-1}}{(0.25)^2 \times 0.01 \times 2 \times 100 \text{ fb}^{-1}} \sim 0.0086$$

where the extra factor of 2 in the denominator corresponds to allowing either of the $g$'s to not be (mis-)tagged in the 3 $b$-tag case. The result is that the $4b$ rates for option (II) are comparable to those for option (I), and that in option (II) the $b\bar{b}gg$ background is substantially smaller than the $b\bar{b}b\bar{b}$ irreducible background. In option (III) the $4b$ and $2b2g$ backgrounds become comparable due to the lower efficiency for $b$ tagging, coupled with a higher rate for mis-tagging. Clearly, $b$-tagging expectations (II) and (III) yield better statistical significance for the signal than the conservative choices of (I), for the same integrated luminosity.

Finally, we recall that at high luminosity it is assumed that 30 GeV is the lowest $p_T^{\min}$ value that can be employed. This effectively restricts the high luminosity option (III) case to $m_{H^0}$ values above about 150 GeV.

In order to compute the statistical significance of a signal when both the $H^0 \to A^0 A^0$ and $H^0 \to h^0 h^0$ decays are present (low $m_{A^0}$) we must take into account the fact that $m_{h^0} \sim m_{A^0}$ at low $m_{A^0}$ once $\tan\beta \gtrsim 4-6$. Our procedure is to include in the $h^0 h^0$ ($A^0 A^0$) signal event rate the portion of the $A^0 A^0$ ($h^0 h^0$) signal event rate that overlaps into the double $[0.88 m_{h^0}, 1.08 m_{h^0}]$ ($[0.88 m_{A^0}, 1.08 m_{A^0}]$) pair mass window. Obviously, when $m_{A^0}$ and $m_{h^0}$ are very close, this simply means the signal rates are added together before the statistical significance $N_{SD} = S/\sqrt{B}$ is computed. Our procedure implies that the contours for the $h^0 h^0$ and $A^0 A^0$ modes will converge for $\tan\beta \gtrsim 4-6$.

We display in Fig. 2 the $N_{SD} = 2, 5,$ and 10 discovery contours for $H^0 \to h^0 h^0$ and $H^0 \to A^0 A^0$ for the $L = 30$ fb$^{-1}$ and $L = 300$ fb$^{-1}$ $b$-tagging options (II) (window 1) and (III) (window 2). It turns out that the results for $b$-tagging option (I) for $L = 100$ fb$^{-1}$ are almost identical to those for option (II) at $L = 30$ fb$^{-1}$ and so we do not present them here.

In the $L = 30$ fb$^{-1}$ option (II) case, we see two distinct regions. At low $m_{A^0}$ values there is a narrow band in the $20 \text{ GeV} \lesssim m_{A^0} \lesssim 60$ GeV region for which $N_{SD} \geq 5$ in both the $h^0 h^0$ and $A^0 A^0$ modes. (The $h^0 h^0$ mode is viable for slightly more parameter space as indicated by its contours which bulge a bit beyond those for the $A^0 A^0$ mode at lower $\tan\beta \lesssim 6$ values.) The lower $m_{A^0}$ limits of these contours are an artifact of our zeroing all rates when the light Higgs mass is below 20 GeV — in any case, we already know from LEP I that $m_{A^0} \geq 40$ GeV. At higher $m_{A^0}$ values there is a $N_{SD} \geq 5$ region for the $h^0 h^0$ mode, extending from roughly $m_{A^0} \sim 100$ GeV up to $m_{A^0} \gtrsim 2 m_t$ for $\tan\beta \lesssim 4$. Note that a viable signal is maintained for $m_{H^0}$ somewhat beyond $2 m_t$, despite the fact that $H^0 \to t\bar{t}$ decays rapidly become dominant. This is because the top-quark loop contribution to the $gg \to H^0$ coupling is substantially enhanced in the vicinity of $m_{H^0} \sim 2 m_t$, thereby enhancing the $gg \to H^0$ production rate.

In the $L = 300$ fb$^{-1}$ option (III) case, the bands at low $m_{A^0}$ disappear; the $p_T > 30$ GeV cut eliminates them by virtue of kinematics. The $N_{SD} \geq 5$ region for $H^0 \to h^0 h^0 \to 4b$ at large $m_{A^0}$ extends to $\tan\beta \lesssim 5$ and slightly higher $m_{A^0}$ values.

### III. TEVATRON RESULTS

In this section, we explore the capability of the Tevatron to observe the $H^0 \to h^0 h^0$ and $H^0 \to A^0 A^0$ signals in the $4b$ final state. CDF and D0 have dramatically increased their $b$-tagging efficiency and purity. At an upgraded Tevatron, they now expect to achieve $e_{b-tag} \sim 0.5$ with $e_{mis-tag} \sim 0.005$ for a $b$-jet with $p_T \gtrsim 15$ GeV and $|\eta| < 2$ at instantaneous luminosities capable of yielding $L = 10 - 30$ fb$^{-1}$ per year [8]. Their ability to obtain such a very small mis-tagging probability is crucial at the Tevatron, since the low event rates make it necessary to accept events in which either 3 or 4 $b$'s are tagged.

Our results for $L = 30$ fb$^{-1}$ are displayed in Fig. 3. For the above-stated $b$-tagging efficiency and purity (left window of figure), a $N_{SD} = 4$ ($N_{SD} = 2$) signal is obtained for $33 \lesssim m_{A^0} \lesssim 53$ GeV ($25 \lesssim m_{A^0} \lesssim 55$ GeV) in both the $h^0 h^0$ and $A^0 A^0$ modes (with the $h^0 h^0$ mode contours again bulging a bit beyond the $A^0 A^0$ contours for $\tan\beta \lesssim 6$). If the detector design can be further improved, so as to achieve $e_{b-tag} \sim 0.75$ and $e_{mis-tag} \sim 0.0025$, the $h^0 h^0$ and $A^0 A^0$ signals will be promoted to at least the $N_{SD} = 5$ level (inner contours of the right-hand window) for most of this low $m_{A^0}$ band. The region for which $H^0 \to h^0 h^0$ is observable at higher $m_{A^0}$ is much more restricted than at the LHC.

We note that the near degeneracy of the $A^0$ and $h^0$ in the low-$m_{A^0}$ region is quite critical to there being an observable level for the $H^0$ signal at the Tevatron. After cuts, tagging, and mass binning, the $4b$ background is typically of order fifty to two hundred events and the $2b2g$ background ranges from several hundred to more than a thousand events. The $A^0 A^0$ and $h^0 h^0$ signals, if not simultaneously present, would typically be just below the observable level; but in combination, they provide a just adequate event rate.

## IV. DISCUSSION AND CONCLUSIONS

We have demonstrated that detection of $gg \to H^0 \to h^0 h^0$ in $4b$ final states at the LHC will be possible for a substantial portion of parameter space, for anticipated integrated luminosities and $b$-tagging efficiency and purity. Should $m_{A^0}$ turn out to be small ($\lesssim 60$ GeV), then *both* $H^0 \to h^0 h^0 \to 4b$ and $H^0 \to A^0 A^0 \to 4b$ yield viable signals. These channels add to the growing list of signals that can provide probes of the MSSM Higgs sector at the LHC [1]. Detection of the $H^0$ in the $h^0 h^0$ and $A^0 A^0$ modes at the Tevatron is mostly limited to the $m_{A^0} \lesssim 55$ GeV portion of parameter space, and requires accumulated luminosity of $L \sim 30$ fb$^{-1}$ as well as excellent $b$-tagging efficiency and purity.

Although observation of $e^+ e^- \to h^0 A^0$ production at LEP II would also be possible in the low

$m_{A^0}$ region of parameter space, the information gained about the Higgs sector from detection at the Tevatron (or LHC) of $gg \to H^0 \to h^0 h^0, A^0 A^0$ would be very complementary.

In our analysis, we have assumed that it will be possible to trigger on the $4b$ final states, perhaps by requiring 4 jets with $p_T > 15$ to 20 GeV, or via a fast $b$-vertex trigger, or some combination of these approaches. This issue is under study by the ATLAS and CMS collaborations.

We note that our computations have not included higher-order QCD $K$ factors for either signal or backgrounds. The $K$ factor for $gg \to H^0$ is known to be substantial ($K \sim 1.6$ [9]) and presumably those for the backgrounds will also be significant. Assuming that all $K$ factors are of similar size (if anything the $gg \to H^0$ $K$ factor is likely to be the largest), the quoted statistical significances for observation will be increased by a factor of $\sqrt{K}$.

The rate for $gg \to H^0 \to h^0 h^0, A^0 A^0$ probes the product, $\Gamma(H^0 \to gg) \times BR(H^0 \to h^0 h^0, A^0 A^0)$, which is sensitive to the $H^0 \to gg$ one-loop coupling and the tri-linear Higgs self couplings. $\Gamma(H^0 \to gg)$ can deviate significantly from the values employed here if squarks are light, or if there are other unobserved heavy colored particles with significant couplings to the $H^0$. The tri-linear Higgs couplings could also deviate from MSSM predictions if, for example, Higgs singlet fields are added to the minimal two-doublet structure. In the very substantial portion of parameter space where both $H^0 \to h^0 h^0 \to 4b$ and $H^0 \to h^0 h^0 \to 2b2\gamma$ can be observed at the LHC, the ratio of the $h^0 \to \gamma\gamma$ and $h^0 \to b\bar{b}$ couplings can be extracted. Clearly, much important information regarding the SUSY Higgs sector, and many important checks of the MSSM, will be made possible by observation of the $H^0 \to h^0 h^0$ and/or $A^0 A^0$ modes.

In Ref. [6], we demonstrated that detection of $gg \to b\bar{b}H^0$ and/or $gg \to b\bar{b}A^0$ in $4b$ final states is possible for sufficiently large $\tan\beta$ values (the required $\tan\beta$ value increases as $m_{A^0}$ increases). Updated results that incorporate the current more optimistic expectations for $b$-tagging efficiency and purity will appear shortly [10]. The region of the $(m_{A^0}, \tan\beta)$ parameter space for which the above modes are observable is largely complementary to that for which we have demonstrated viability for the $4b$ final states resulting from $gg \to H^0 \to h^0 h^0$ and/or $gg \to H^0 \to A^0 A^0$. In combination, these two different types of $4b$ modes allow $H^0$ and/or $A^0$ detection over a very sizeable fraction of the $(m_{A^0}, \tan\beta)$ parameter space plane when $m_{A^0} \lesssim 400$ GeV. There is a wedge between the discovery regions at moderate $\tan\beta \gtrsim 5$ that grows as $m_{A^0}$ increases above 400 GeV. Further improvements in $b$-tagging efficiency and purity would substantially reduce this wedge. Efforts by the ATLAS and CMS collaborations in this direction are thus highly desirable. Overall, the $4b$ modes have considerably enhanced the prospects for detecting the $H^0$ and/or the $A^0$ at the LHC. It is thus quite possible that the LHC will be able to fully explore the Higgs sector of the MSSM.

## ACKNOWLEDGMENTS

This work was supported in part by the U.S. Department of Energy. Further support was provided by the Davis Institute for High Energy Physics.

**FIGURES**

1. We show contours of fixed $BR(H^0 \to h^0 h^0) = 0.5$, 0.2, 0.1 and 0.05 in the $(m_{A^0}, \tan\beta)$ parameter space. We have taken $m_t = 175$ GeV (pole mass) and $m_{\tilde{t}} = 1$ TeV, and neglected squark mixing. The 0.2, 0.1 and 0.05 contours in the vicinity of $m_{A^0} \sim 60$ GeV are essentially indistinguishable; the $H^0 \to h^0 h^0$ decay becomes kinematically disallowed for $m_{A^0}$ values just beyond this boundary.

2. We show the $(m_{A^0}, \tan\beta)$ parameter space contours within which $H^0 \to h^0 h^0 \to 4b$ and $H^0 \to A^0 A^0 \to 4b$ can be observed at the 2, 5, or $10\sigma$ level assuming: 1) an integrated luminosity of $L = 30$ fb$^{-1}$ and 4-$b$-tagging option (II); and 2) $L = 300$ fb$^{-1}$ and 4-$b$-tagging option (III). See text for details. We take $m_t = 175$ GeV and include two-loop radiative corrections, assuming $m_{\tilde{t}} = 1$ TeV and neglecting squark mixing. We also assume that SUSY decays are absent.

3. We show the $(m_{A^0}, \tan\beta)$ parameter space contours within which $H^0 \to h^0 h^0, A^0 A^0 \to 4b$ can be observed: 1) at the 2, 4 $\sigma$ level assuming $\epsilon_{b-tag} = 0.5$ and $\epsilon_{mis-tag} = 0.005$; and 2) at the 3, 5 $\sigma$ level assuming $\epsilon_{b-tag} = 0.75$ and $\epsilon_{mis-tag} = 0.0025$. We assume $\sqrt{s} = 1.8$ TeV, an integrated luminosity of $L = 30$ fb$^{-1}$ and accept events in which either 3 or 4 $b$'s with $p_T > 15$ GeV and $|\eta| < 2$ are tagged.

FIGURES

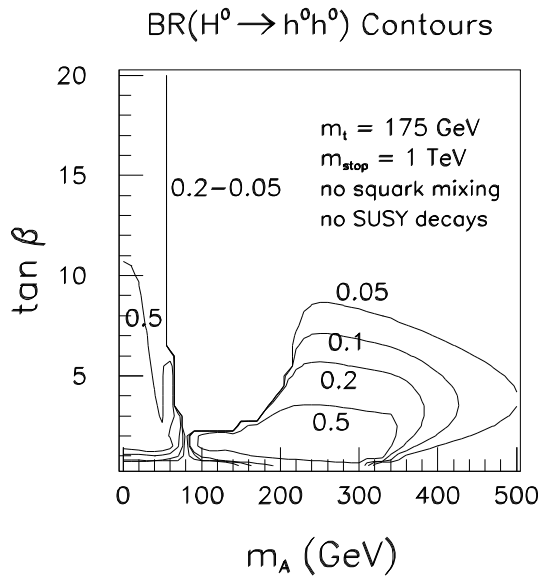

FIG. 1. We show contours of fixed $BR(H^0 \to h^0 h^0) = 0.5, 0.2, 0.1$ and 0.05 in the $(m_{A^0}, \tan\beta)$ parameter space. We have taken $m_t = 175$ GeV (pole mass) and $m_{\tilde{t}} = 1$ TeV, and neglected squark mixing. The 0.2, 0.1 and 0.05 contours in the vicinity of $m_{A^0} \sim 60$ GeV are essentially indistinguishable; the $H^0 \to h^0 h^0$ decay becomes kinematically disallowed for $m_{A^0}$ values just beyond this boundary.

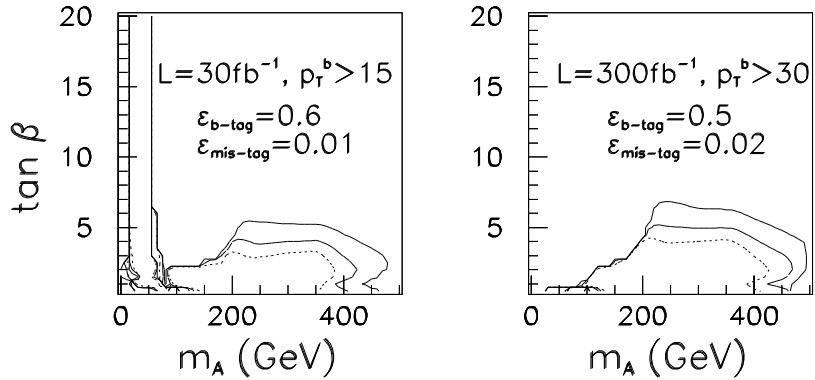

FIG. 2. We show the $(m_{A^0}, \tan\beta)$ parameter space contours within which $H^0 \to h^0 h^0, A^0 A^0 \to 4b$ can be observed at the 2, 5, or $10\sigma$ level assuming: 1) an integrated luminosity of $L = 30$ fb$^{-1}$ and 4-$b$-tagging option (II); and 2) $L = 300$ fb$^{-1}$ and 4-$b$-tagging option (III). See text for details.

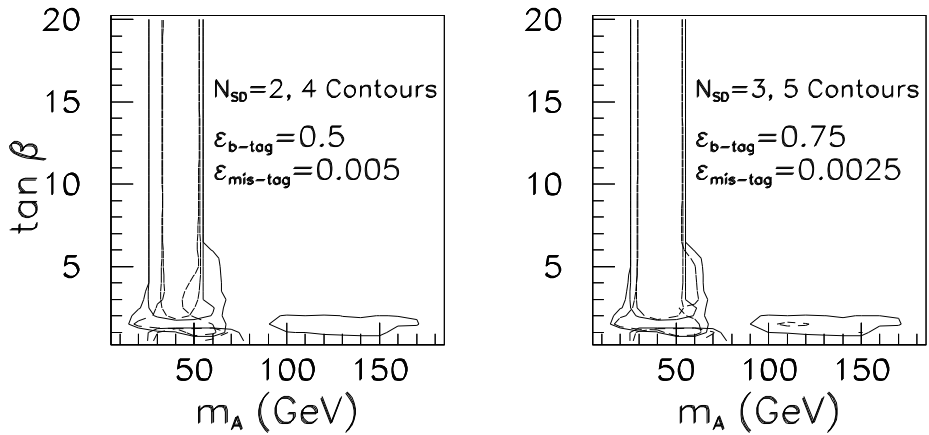

FIG. 3. We show the $(m_{A^0}, \tan\beta)$ parameter space contours within which $H^0 \to h^0 h^0, A^0 A^0 \to 4b$ can be observed: 1) at the 2, 4 $\sigma$ level assuming $e_{b-tag} = 0.5$ and $e_{mis-tag} = 0.005$; and 2) at the 3, 5 $\sigma$ level assuming $e_{b-tag} = 0.75$ and $e_{mis-tag} = 0.0025$. We assume $\sqrt{s} = 1.8$ TeV, an integrated luminosity of $L = 30$ fb$^{-1}$ and accept events in which either 3 or 4 $b$'s with $p_T > 15$ GeV and $|\eta| < 2$ are tagged.